\documentstyle[12pt]{article}
\begin{document}
\date{\today}
\pagestyle{plain}
\title{The Space-time Transformations Between Accelerated Systems}
\author{Miroslav Pardy \\[2cm]
Institute of Plasma Physics ASCR\\
Prague Asterix Laser System, PALS\\
Za Slovankou 3, 182 21 Prague 8, Czech Republic\\
and\\
Department of Physical Electronics \\
Masaryk University \\
Kotl\'{a}\v{r}sk\'{a} 2, 611 37 Brno, Czech Republic\\
e-mail:pamir@physics.muni.cz}
\date{\today}
\maketitle
\vspace{50mm}

\begin{abstract}
We determine transformations between coordinate systems which are mutually
in linear accelerated motion. We also determine the transformations
for rotating systems.
In case of the symmetrical linear mutual acceleration, we immediately get the
maximal acceleration limit which was derived by Caianiello
from quantum mechanics.
Maximal acceleration is an analogue of maximal velocity in special relativity.
We discuss the possible verification of derived formulae by the measurement
with ultracentrifuge. It is argued that the derived
results can play crucial role in modern particle physics and cosmology.
\end{abstract}

\newpage

\baselineskip 15 pt

\section{Introduction}

The problem of acceleration of charged particles or systems of particles
is the permanent and the most prestige problem in the accelerator
physics. Particles can be accelerated by different ways. Usually by
the classical electromagnetic fields, or, by light pressure
of the laser fields (Ashkin, 1970, 1972; Baranova et al. 1994;
Pardy, 1998, 2001, 2002).
The latter method is the permanent problem of the laser physics for
many years.

However, the theoretical problem is not only to find the mechanisms of
acceleration bat also the space-time relations between systems which are
mutually in the accelerated motion. The uniformly accelerated systems
are well known, also the rotating systems represented for instance by the
centrifuges are also known and theoretically investigated
by many authors.

Here, we determine transformations between coordinate systems
which moves mutually with acceleration. We determine transformations
between nonrelativistic and relativistic uniformly accelerated systems
and rotating systems. We derive also
some consequences following from the nonlinearity of motion of these systems.

We show that the transformation laws between accelerated systems can
be derived  from the infinitesimal Lorentz transformation on the one
hand, or, by postulation some kinematical symmetries between these
systems on the other hand. These two approaches gives
different results. In the case of the first approach
which is based on the original
Lorentz transformation, the derived results can be taken for sure-footed.

We do not consider in this article the problem of accelerated strings,
which is solved for instance by Bachas (2002),
because string theory is under the permanent reconstruction and according to
Witten (Witten, 2002) it is not clear that the present form of the string
theory is correct.

\section{The infinitesimal form of the Lorentz transformation}

We know, that the Lorentz transformation between two inertial coordinate
systems $S(0, x, y, z)$ and $S'(0, x', y', z')$ (where system $S'$ moves in
such a way that $x$-axes converge, while $y$ and $z$-axes run parallel and at
time $t = t' = 0$ for the beginning of the systems $O$ and $O'$ it is $O
\equiv O'$) is as follows:

$$x' = \gamma(v)(x - vt), \quad y' = y,\quad z' = z',
\quad t' = \gamma(v)\left(t - \frac
{v}{c^{2}}x\right),\eqno(1)$$
where

$$\gamma(v) = \left(1 - \frac {v^{2}}{c^{2}}\right)^{-1/2}.\eqno(2)$$

The infinitesimal form of this transformation is evidently given by
differentiation of the every equation. Or,

$$dx' = \gamma(v)(dx - vdt), \quad dy' = dy,\quad dz' = dz,
\quad dt' = \gamma(v)\left(dt - \frac{v}{c^{2}}dx\right).\eqno(3)$$

If we put $dt = 0$ in the first equation of system (3), then
the Lorentz length contraction follows
in the infinitesimal form $dx' = \gamma(v)dx$. Or, in other words,
if in the system $S'$
the infinitesimal length is $dx'$, then the relative length with
regard to the system
$S$ is $\gamma^{-1}dx'$. Similarly, from the
last equation of (3) it follows the time dilatation for $dx = 0$.
Historical view on this effect is in the Selleri (1997) article.

If the velocity depends on time, which is for instance in the case of the
nonlinear motion, then we write

$$dx' = \gamma(v(t))(dx - v(t)dt), \quad dy' = dy,\quad dz' = dz,
\quad dt' = \gamma(v(t))\left(dt - \frac{v(t)}{c^{2}}dx\right).\eqno(4)$$

This infinitesimal form enables the integration and if we know the dependence
of $v$ on time, then there is no obstacles to get the Lorentz-like
transformation between two nonlinear systems. At the same time the
transformations (4) does not change the so called Minkowski metric element,
the square of which is

$$ds^{2} = c^{2}dt^{2} - dx^{2} -dy^{2} - dz^{2}.\eqno(5)$$

The Lorentz transformation for coordinates and time referring to two inertial
systems harbours the assumption that the following expression holds good:
$(x = ct)\quad \Longleftrightarrow \quad (x' = ct')$,
where the invariant function $x = ct$ being considered in the special theory
of relativity as the mathematical expression of the principle of constant
light velocity. From the mathematical point of view the relation is the
formal mathematical requirement for unambiguous determination of the Lorentz
transformation and it follows from the theory of the continuous group of
transformations (Eisenhart, 1943).
The physical meaning of $ds$ is usually defined as a distance between
two infinitesimal events, however, some
authors consider $ds$  as a formal mathematical object
with no physical meaning (Brillouin, 1970).

\section{Uniformly accelerated systems}

According to Einstein (1965), Fok (1961) and Logunov (1987), space and time,
or, space-time is a form of the existence of matter.
To study space-time means to study
form of the existence of matter. Special theory of relativity
investigates behavior of
matter and fields in case of the inertial systems in the inertial motion.
The behavior of space-time in case that
the systems are mutually or individually accelerated is investigated here.

Let us suppose that in the finite time interval the system $S'$ is accelerated
by the constant acceleration in such a way that the motion is nonrelativistic
one. So, for the velocity of the system $S'$ we have $v = at$ and from
the equations (4) we have for the infinitesimal Lorentz transformation:

$$dx' = \gamma(at)(dx - atdt), \quad dy' = dy,\quad dz' = dz,
\quad dt' = \gamma(at)\left(dt - \frac{at}{c^{2}}dx\right),\eqno(6)$$
where

$$\gamma(at) = \left(1 - \frac {a^{2}t^{2}}{c^{2}}\right)^{-1/2}.\eqno(7)$$

After integration of equation (6) we get for the coordinate
and time components:

$$x' = x\left(1 - \frac {a^{2}t^{2}}{c^{2}}\right)^{-1/2} -
a\int t dt\left(1 - \frac {a^{2}t^{2}}{c^{2}}\right)^{-1/2} = $$

$$
x\left(1 - \frac {a^{2}t^{2}}{c^{2}}\right)^{-1/2} +
\frac {c^{2}}{a}\left(1  - \frac {a^{2}t^{2}}{c^{2}}\right)^{1/2}
\eqno(8)$$
and

$$t' =
\frac {c}{a}\arcsin \left(\frac {at}{c}\right) -
x\frac {at}{c^{2}}\left(1 - \frac {a^{2}t^{2}}{c^{2}}\right)^{-1/2}.
\eqno(9)$$

In case of the relativistic motion of a body with mass $m$ which is caused
by the action of the constant force $F$ on this body,
the dependence of velocity on time is (Landau et al. 1962)

$$v = \frac {at}{\sqrt{1 + \left(\frac {at}{c}\right)^{2}}}.\eqno(10)$$

Then, the relativistic coefficient $\gamma(v)$ is given by the relation

$$\gamma(v) = \left(1 + \frac {a^{2}t^{2}}{c^{2}}\right)^{1/2}.\eqno(11)$$

In this situation we get for the coordinate and time transformation:

$$x' = x\left(1 + \frac {a^{2}t^{2}}{c^{2}}\right)^{1/2} -
a\int t dt =
x\left(1 + \frac {a^{2}t^{2}}{c^{2}}\right)^{1/2} -
\frac {a}{2}t^{2}
\eqno(12)$$
and

$$t' =
\frac {1}{2ac}\left[at\sqrt{a^{2}t^{2} + c^{2}} +
c^{2}\arg\sinh\left(\frac {at}{c}\right)\right] -
x\frac {at}{c^{2}},
\eqno(13)$$
where $\arg\sinh(at/c)$ can be expressed in the logarithmic form according
to he following formula:

$$\arg\sinh\left(\frac {at}{c}\right) = \ln\left[
\left(\frac {at}{c}\right) +
\sqrt{\left(\frac {at}{c}\right)^{2} + 1}\right].\eqno(14)$$

Let us remark, that the problem of transformations between nonlinear systems
was also discussed by M\o ller (1943)  and Fok (1961). The
transformations derived by M\o ller are not identical with ones derived by
us. According to Fok, the nonlinear transformation of space-time  is as
follows:

$$x' = x\cosh\left(\frac {at}{c}\right) -
\frac{c^{2}}{a}\left(\cosh\left(\frac {at}{c}\right) -1\right)\eqno(15a)$$
and

$$t' = \frac {c}{a}\sinh\left(\frac {at}{c}\right) -
\frac
{x}{c}\sinh\left(\frac {at}{c}\right).\eqno(15b)$$

The trasnformations between system $S$ and uniformly accelerated system $S'$
was also derived by  Logunov (1987), but in the substantially different form:

$$x' = x \frac {c^{2}}{a}\left[\sqrt{1 + \frac {a^{2}t^{2}}{c^{2}}} -1\right];
\quad t' = t.\eqno(16)$$
where equation $t' = t$ can be chosen according to Logunov as the free
decision of a physicist. The Logunov approach is not in the final form.

The transformations derived reduces  for

$$\frac {at}{c} \ll  1\eqno(17)$$
to the Galilean transformation in the form

$$x' = x - \frac {1}{2}at^{2}; \quad t' =  t.\eqno(18)$$

After insertion of transformation (15) into space-time element
(4), we get:

$$ds^{2} = \left(c - \frac {ax}{c}\right)^{2}dt^{2}
 - dx^{2} - dy^{2} - dz^{2}\eqno(19)$$
which is approximately (for $|ax| \ll c^{2}$) the same as the square of
the space-time element in
the homogenous gravitational potential $U = ax$.
It means that the uniformly accelerated system form some analogue with the
homogenous gravitational field. The analogy is usually defined as the
principle of equivalence.

However, the principle of equivalence can be easily derived from the special
theory relativity from the well known relation $E = mc^{2}$.
 This relation means
that to every mass corresponds energy. And the energy has the same inertion
in the arbitrary acceleration field. It means, there is no difference
between gravitational and inertional mass. So, the principle of equivalence
between inertial and gravitational mass follows from the Einstein relation
$E = mc^{2}$. From this proof it follows, that no future experiment will
reveal the difference between inertial and gravitational mass. To our
knowledge, this elementary theorem was not published in any journal.

We can say, that there is some different nonequivalent possibilities
in the derivation of the
transformations for nonlinearly moving systems. It signalizes that
the theory
of the space-time transformation between nonlinear systems
is not in the definite form.

The inverse transformations to the derived ones are evidently of the
different form (excepting the Logunov transformation)
than the original transformations.
We show, In the following section that it is possible to find such
transformations between coordinates and time, that they are symmetrical.
The physical meaning of such transformation is open.

\section{Uniformly accelerated frames with space-time symmetry}

Let us take two systems $S(0, x, y, z)$ and $S'(0, x', y', z')$,
where system $S'$ moves in
such a way that $x$-axes converge, while $y$ and $z$-axes run parallel and at
time $t = t' = 0$ for the beginning of the systems $O$ and $O'$ it is $O
\equiv O'$. Let us suppose that system $S'$ moves relative to some basic
system $B$
with acceleration $a/2$ and system $S'$ moves relative to system $B$
with acceleration $-a/2$. It means that both systems moves one another
with acceleration $a$ and are equivalent because
in every system it is possibly to observe the force caused by the
acceleration $a/2$. In other word no system is inertial.

Now, let us consider the formal transformation equations between two systems.
At this moment we give to this transform only formal meaning because at this
time, the physical meaning of such transformation is not known.
On the other hand,
the consequences of the transformation will be shown very interesting.

We write the transformation equations in the form:

$$x' = a_{1}x + a_{2}t^{2}, \quad y' = y,\quad z' = z,
\quad t' = \sqrt{b_{1}x + b_{2}t^{2}},\eqno(20)$$
where constants involved in the equations will be determined from the
viewpoint of kinematics. Since from the viewpoint of kinematics, both systems
are equivalent, for the inverse transformation to the transformation (20)
it must hold:

$$x = a_{1}x' - a_{2}t'^{2}, \quad y' = y,\quad z' = z,
\quad t = \sqrt{-b_{1}x' + b_{2}t'^{2}}.\eqno(21)$$

The minus sign with coefficients $a_{2}$ and  $b_{1}$ appearing for the reason
that constant $a_{2}$ has the rate of acceleration while constant $b_{1}$
the rate of inverse value of acceleration.

Similarly as in inertial systems, the hypothetical requirement can be now
expressed that the transformation equations for system moving relative to
themselves with acceleration include a suitable invariant function. Let us now
define such transformations as follows:

$$x = \frac {1}{2}\alpha t^{2},\eqno(22a)$$

$$x' = \frac {1}{2}\alpha t'^{2},\eqno(22b)$$
where $\alpha$ is the constant having the rate of acceleration.

If we now substitute (21) into (20)
we obtain

$$x' = x'(a_{1}^{2}- a_{2}b_{1}) + t'^{2}(a_{2}b_{2} -
a_{1}a_{2}),\eqno(23)$$

$$t'^{2} = x'(a_{1}b_{1} - b_{1}b_{2}) + t'^{2}(b_{2}^{2} -
b_{1}a_{2}).\eqno(24)$$

After comparing the left and right sides in the relations (23), we get

$$a_{1} = b_{2}, \quad a_{1}^{2} -  b_{1}a_{2} = 1. \eqno(25)$$

If we put in the relation (20) $x'$ = 0, we obtain $x = -(a_{2}/a_{1})t^{2}$.
In accordance with the assumption the motion of the beginning of the system
$S'$ relative to system $S$ is described by the
$x = \frac{1}{2}at^{2}$, we thus obtain

$$a_{2} = - \frac {1}{2}a_{1}a. \eqno(26).$$

From $[x' = (1/2)\alpha t'^{2}] \Longleftrightarrow
[x = (1/2)\alpha t^{2}$], we get

$$\frac{
\frac{1}{2}\alpha b_{2} - a_{2}}{a_{1} - \frac {1}{2}\alpha b_{1}} =
\frac {1}{2}\alpha.\eqno(27)$$

Through solving the equations (26) and  (27), we obtain

$$a_{1} = b_{2} = \frac {1}{\sqrt{1 - \frac {a^{2}}{\alpha^{2}}}},\quad
a_{2} = -\frac {1}{2}\frac {a}{\sqrt{1 - \frac {a^{2}}{\alpha^{2}}}},\quad
b_{1} = -\frac {2}{\alpha^{2}}\frac {a}{\sqrt{1 - \frac {a^{2}}
{\alpha^{2}}}}.\eqno(28)$$

Using (28), we can rewrite the transformation (20) in the definite form:

$$x' = \Gamma(a)(x - \frac {1}{2}at^{2}), \quad y' = y,\quad z' = z,
\quad t'^{2} = \Gamma(a)\left(t^{2} - \frac{2a}{\alpha^{2}}x\right)
\eqno(29)$$
with

$$\Gamma(a) = \frac {1}{\sqrt{1 - \frac {a^{2}}{\alpha^{2}}}}.\eqno(30)$$

Let us remark that the more simple derivation of the last transformation can be
obtained if we perform in the Lorentz transformation the elementary
change of variables as follows:
$t \rightarrow t^{2},\quad t' \rightarrow t'^{2}, \quad  v \rightarrow
\frac {1}{2}a, \quad c \rightarrow \frac {1}{2}\alpha$.

After performing such elementary transition which is practically
redenotation
of variables, we really get the Lorentz-like transformation  (29) between
accelerated systems.

The physical interpretation of this nonlinear transformations is the
same as in the case of the Lorentz transformation only the physical
interpretation of the invariant function $x = (1/2)\alpha t^{2}$ is
open.

However, we know from history, that Lorentz
transformation was taken first as physically meaningless by Lorentz
himself and later only Einstein decided to put the physical meaning
to this transformation and to the invariant function $x = ct$.
We hope that the derived transformation will appear as physically
meaningful.

Now, let us prove the following assumption. The transformation (29) forms
one-parametric group with parameter $a$. To prove it we must prove
by the direct calculations the four requirements
involving in the definition of group. However, we know,
that using relations $t \rightarrow t^{2},\quad t' \rightarrow t'^{2},
 \quad  v \rightarrow
\frac {1}{2}a, \quad c \rightarrow \frac {1}{2}\alpha$, the nonlinear
transformation is expressed as the Lorentz transformation
forming the one-parametric group. And this is a proof.
Such proof is equivalent to the proof by direct calculation.
The integral part of the group properties is the so called
addition theorem for acceleration.

$$w_{3} = \frac {w_{1} + w_{2}}{1 + \frac {w_{1}w_{2}}{\alpha^{2}}}.
\eqno(31)$$
where $w_{1}$ is the acceleration of the $S'$ with regard to the system $S$,
$w_{2}$ is the acceleration of the system $S''$ with regard to the system
$S'$ and $w_{3}$ is the acceleration of the system $S''$  with regard to the
system $S$.

The relation (31),
expresses the law of acceleration addition theorem
on the understanding that the
events are marked according to the relation  (29). In this formula as well as
in the transformation equation (29) appears constant $\alpha$ which cannot be
calculated from the theoretical considerations, or, from the theory.
What is its magnitude and whether there exists such a physical field that is
consistent with the designation of the events given by the relations (29) can
be established only by  experiments. On the other hand the constant
$\alpha$ has physical meaning of the maximal acceleration and its meaning
is similar to the maximal velocity $c$ in special relativity. The notion
maximal acceleration is not new
physics, because Caianiello (1981, 1992) introduced it as some consequence
of quantum mechanics and Landau theory of fluctuations. Revisiting view
on the maximal acceleration was given by Papini (2003).
At resent time it was
argued by Lambiase et al. (1998, 1999) that
maximal acceleration determines the upper limit
of the Higgs boson and that it gives also the relation which
links the mass of $W$ boson with the mass of the Higgs boson.
The LHC Experiments probably give the answer to this problem.

\section{Dependence of mass on acceleration}

If the maximal acceleration is the physical reality, then it should have the
similar consequences in a dynamics as the maximal velocity of motion has
consequences in the dependence of mass on velocity.
We can suppose in analogy with the special relativity
that mass depends on the acceleration for small
velocities,
in the similar way as it depends on velocity in case of uniform motion.
Of course such assumption must be experimentally verified and in no case
it follows from special theory of relativity, or, general theory of
relativity (Okun, 2001). So, we postulate
ad hoc, in analogy with special theory of relativity:

$$m(a) = \frac {m_{0}}{\sqrt{1 - \frac {a^{2}}{\alpha^{2}}}};
\quad v \ll c, \quad a = \frac{dv}{dt}.\eqno(32)$$

Let us derive as an example the law of
motion when the constant force $F$  acts on the body with the rest
mass $m_{0}$. Then, the Newton law reads (Landau et al., 1962):

$$F = \frac{dp}{dt} =
m_{0}\frac{d}{dt}\frac {v}{\sqrt{1 - \frac {a^{2}}{\alpha^{2}}}}.
\eqno(33)$$

The first integral of this equation can be written in the form:

$$\frac{Ft}{m_{0}} = \frac {v}{\sqrt{1 - \frac {a^{2}}{\alpha^{2}}}};
\quad a = \frac{dv}{dt}, \quad F = const.. \eqno(34)$$

Let us introduce quantities

$$v = y,\quad a = y', \quad A(t) =
\frac{F^{2}t^{2}}{m_{0}^{2}\alpha^{2}}.\eqno(35)$$

Then, the quadratic form of the equation (34) can be written as
the following differential equation:

$$A(t)y'^{2} + y^{2} - A(t)\alpha^{2} = 0, \eqno(36)$$
which is nonlinear differential equation of the first order.
The solution of it is of the form $y = Dt$, where $D$ is some constant,
which can be easily determined. Then, we have the solution in the form:

$$y = v = \frac {t}{\sqrt{\frac {m_{0}^{2}}{F^{2}} + \frac {1}{\alpha^{2}}}}.
\eqno(37)$$

For F $\rightarrow \infty$, we get $v = \alpha^{2}t$. This relation can
play substantial role at the beginning of the big-bang, where the accelerating
forces can be considered as infinite, however the law of acceleration
has finite nonsingular form.
At this moment it is not clear if the dependence of the mass on acceleration
can be related to the energy dependence on acceleration similarly to the
Einstein relation uniting energy, mass and velocity
(Sachs, 1973; Okun, 2001).

\section{The rotating systems}

According to Einstein, there is an analogy between gravitational
fields and noninertial reference
system. Therefore, when studying properties of
gravitational fields in relativistic mechanics, we can start from this
analogy.

The description of the rotation system can be described in the Cartesian
coordinate system (Landau et al., 1962), or, in the more appropriate form
using the polar coordinate $r, \varphi$ (Goy and Selleri, 1997). Then, we write

$$x = r\cos(\varphi + \omega t),\quad y = r\sin(\varphi + \omega t).
\eqno(38)$$

The corresponding space-time element is as follows:

$$ds^{2} = \left(1 - \frac {\omega^{2}r^{2}}{c^{2}}\right)\left(cdt\right)^{2}
- \frac {2\omega r^{2}}{c}d\varphi(cdt) - dz^{2} - dr^{2} - r^{2}d\varphi^{2}.
 \eqno(39)$$

We see from the time term in (39)
that if we suppose that the velocity of light in the
rotating system is constant, then the elapsing of time depends on acceleration.

The equations (38) does not involve the transformation between $t'$  and $t$,
so we have motivation for find such transformation.

Such transformation between inertial and rotating system can be expressed
in the Lorentz form if we insert into the original Lorentz transformation
the following formulae $\Delta x = \Delta\varphi r, v = \omega r$. Then we have
for $0 \le \varphi \le 2\pi$:

$$\Delta\varphi' = \frac {\Delta\varphi - \omega \Delta t}{\sqrt{1 - \frac
{\omega^{2}r^{2}}{c^{2}}}},\quad
\Delta t' = \frac {\Delta t - \Delta\varphi \frac {\omega r^{2}}{c^{2}}}
{\sqrt{1 - \frac {\omega^{2}r^{2}}{c^{2}}}}.\eqno(40))$$

It follows from this transformation that for every $r > c/\omega$
and given $\omega$,
it has no physical meaning. It also means that the infinite
rotating system as a whole has no relativistic meaning
and it is not clear how to solve this problem.
We also see that from equation (40) the time dilatation follows and it
is the same as in equation (39).

Although the rotating system cannot be considered as
equivalent to the linear accelerated system, nevertheless,
the radial component of every part of this system is in the
permanent acceleration.

If the element 1 of the rotating plane at the radial coordinate $r_{1}$ has
acceleration $w_{1}$  and
if the element of the rotating plane at the radial coordinate $r_{2}$ has
acceleration $w_{2}$,
then the relative acceleration $w_{r}$ of the element 2 with regard to
the element 1 is not $w_{2} - w_{1}$, but must be determined
according to the formula

$$w_{r} = \frac {w_{2} - w_{1}}{1 - \frac {w_{1}w_{2}}{\alpha^{2}}}.
\eqno(41)$$

The last formula is an analogue of the formula which determines the
relative velocities in case of the inertial motion in the special
theory of relativity.
The last formula is true only if the transversal effect do not influence the
radial effects.
It can be verified optically, because we know that the optical frequency of
the emission source is influenced by acceleration, or, equivalently by the
gravitational field.

Similarly, it is possible to verify the dependence of mass
on acceleration, also by the ultracentrifuge.

\section{Discussion}

We have derived transformations
between accelerated systems moving mutually uniformly.
We have discussed also the rotating systems.
We have derived some consequences following from the nonlinearity of
motion of these systems. In case, when we used the symmetry principle
in derivation of the space-time transformation, we derived by the
formal way so called maximal acceleration which was derived
using quantum mechanics by Caianiello (1981, 1992). Our derivation of the  maximal
acceleration is not equivalent to the Caianiello derivation and at the same
time it is not in the contradiction with his approach because the heuristical
ways to the maximal acceleration were substantially different.

If some experiment will confirm the existence of maximal
acceleration $\alpha$, then it will have
certainly crucial consequences for Einstein theory of gravity
because this theory does not involve this factor.
Also the cosmological theories constructed on the basis of the
original Einstein equations will require modifications.
In such a way, Einstein equations can play a role only in the specific
conditions where the maximal acceleration can be neglected.
Maximal acceleration
does not allow the existence of black holes with
arbitrary big mas. Also standard model in particle physics will require
generalization because it does not involve the maximal acceleration.

We did not consider the problem of accelerated strings,
because string theory is under permanent reconstruction and according
to Witten (2002) it is not sure that the present form of this theory
is correct. It also does not involve the Penrose nonlinear graviton
(Penrose, 1976) and it does not involve the Gassendi string
model of gravity (Fraser et al. 1998).

One of the prestige problem in the modern theoretical physics is
the Unruh effect, or, the existence
of thermal radiation detected by accelerated  observer.
The theory of the Unruh effect is unfortunately under the reconstruction
(Fedotov et al., 2002) and to say
serious statement or comment to the relation of this effect
to the maximal acceleration is premature.

So, at this moment, we  study only the accelerated
classical systems, and after some time,
when string theory will be in the perfect form we will study
the situation where the string is accelerated by
quantized laser field, following the author articles
concerning laser acceleration (Pardy, 1998, 2001, 2002).

\vspace{15mm}

{\bf REFERENCES}

\vspace{12mm}

\noindent
Ashkin, A. (1970). Acceleration and trapping of particles by radiation
pressure, {\it Phys. Rev. Lett.} {\bf 24}, 156.
\\[2mm]
Ashkin, A. (1972). The pressure of laser light, {\it Scientific
American} {\bf 226} (2), 63.
\\[2mm]
Bachas, C. (2002). Relativistic string in a pulse, e-print LPTENS 02/63
and hep-th/0212217.\\[2mm]
Baranova, N. B. and  Zel'dovich, B. Ya. (1994). Acceleration of
charged particles by laser beams,  {\it JETP} {\bf 78}, (3), 249.
\\[2mm]
Brillouin, L. (1970). {\it Relativity Reexamined}, Academic Press, New York
and London.\\[2mm]
Caianiello, E. R. (1981). Is there a maximal acceleration ?,
{\it Lett. Nuovo Cimento} {\bf 32}, 65; ibid. (1992).
{\it Revista del Nuovo Cimento}, No. 4. \\[2mm]
Einstein, A. (1965). {\it Collective Scientific Works}, Vol I, Moscow,
(in Russian).
\\[2mm]
Eisenhart, L. P. (1943). {\it Continuous Groups of Transformations},
Princeton.
\\[2mm]
Fedotov, A. M., Narozhny, N. B., Mur, V. D. and Belinski, V. A. (2002).
An example of a uniformly accelerated particle detector with non-Unruh
response,  e-print hep-th/0208061.\\[2mm]
Fok, V. A. (1961). {\it The theory of Space, Time and Gravity},
second edition, GIFML, Moscow, (in Russian).
\\[2mm]
Fraser,G., Lillest\o l, E. and Sellav\aa g, I. (1998).
{\it The Search for Infinity}
George Philip Limited, second ed..
\\[2mm]
Goy, F. and Selleri, F. (1997). Time on a rotating platform,
e-print gr-qc/9702055. \\[2mm]
Lambiase, G, Papini, G. and Scarpetta, G. (1998). Maximal acceleration
limits on the mass of the Higgs boson, e-print hep-ph/9808460 and ibid.
(1999). {\it Nuovo Cimento B}, Vol. {\bf 114}, 189.
\\[2mm]
Landau, L. D. and Lifshitz, E. M. (1962). {\it The Classical
Theory of Fields}, 2nd ed.~ Pergamon Press, Oxford.
\\[2mm]
Logunov, A. A.  (1987). {\it Lectures on the
Theory of Relativity and Gravitation},
Nauka, Moscow, (in Russian).\\[2mm]
M\o ller, C. (1943). On homogenous gravitational
fields in the general theory of
relativity and the clock paradox,  {\it Works of
Denmark Academy of Sciences}, Vol. {\bf 2}, No. 19. Kobenhavn.\\[2mm]
Okun, L. B. (2001). Photons, clocks, gravity and concept of mass,
e-print physics/0111134. \\[2mm]
Papini, G., (2003). Revisiting Caianiello's maximal acceleration,
e-print quant-ph/0301142.\\[2mm]
Pardy, M. (1998). The quantum field theory of laser acceleration,
{\it Phys. Lett. A} {\bf 243}, 223.
\\[2mm]
Pardy, M. (2001). The quantum electrodynamics of laser acceleration,
{\it Radiation Physics and Chemistry} {\bf 61}, 391.\\[2mm]
Pardy, M. (2002). Electron in the ultrashort laser pulse, e-print
hep-ph/0207274. \\[2mm]
Penrose, R. (1976). The nonlinear graviton, {\it General
Relativity and Gravitation},  Vol. {\bf 7}, No. 1., 171.\\[2mm]
Sachs, M. (1973). On the meaning of $E = mc^{2}$, {\it International Journal of
Theoretical Physics}, Vol. {\bf 8}, No. 5, 377. \\[2mm]
Selleri, F. (1997). The relativity principle and the nature of time,
{\it Foundations of Physics}, Vol. {\bf 27}, No. 11, 1527.
\\[2mm]
Witten, E. (2002). Comments on string theory, eprint hep-th/0212247.

\end{document}